\documentclass[10pt,conference]{IEEEtran}
\IEEEoverridecommandlockouts
\usepackage{cite}
\usepackage{amsmath,amssymb,amsfonts}
\usepackage{algorithmic}
\usepackage{graphicx}
\usepackage{textcomp}
\usepackage{xcolor}
\usepackage{booktabs} 

\def\BibTeX{{\rm B\kern-.05em{\sc i\kern-.025em b}\kern-.08em
    \kern-.1667em\lower.7ex\hbox{E}\kern-.125emX}}

\begin{document}

\title{Time-Series Foundation Models for ISP Traffic Forecasting\\
\thanks{This work was supported by NSF CNS Award 2213672.}
}

\author{
\IEEEauthorblockN{Fan Liu, Behrooz Farkiani, Patrick Crowley}
\IEEEauthorblockA{\textit{Washington University in St. Louis} \\
1 Brookings Dr, St. Louis, MO 63130, USA \\
\{fan.liu, b.farkiani, pcrowley\}@wustl.edu}
}

\maketitle

\begin{abstract}
Accurate network-traffic forecasting enables proactive capacity planning and anomaly detection in Internet Service Provider (ISP) networks. Recent advances in time-series foundation models (TSFMs) have demonstrated strong zero-shot and few-shot generalization across diverse domains, yet their effectiveness for computer networking remains unexplored. 
This paper presents a systematic evaluation of a TSFM, IBM’s Tiny Time Mixer (TTM), on the CESNET-TimeSeries24 dataset, a 40-week real-world ISP telemetry corpus. We assess TTM under zero-shot and few-shot settings across multiple forecasting horizons (hours to days), aggregation hierarchies (institutions, subnets, IPs), and temporal resolutions (10-minute and hourly). Results show that TTM achieves consistent accuracy (RMSE 0.026-0.057) and stable $R^2$ scores across horizons and context lengths, outperforming or matching fully trained deep learning baselines such as GRU and LSTM. Inference latency remains under 0.05s per 100 points on a single MacBook Pro using CPU-only computation, confirming deployability without dedicated GPU or MPS acceleration. These findings highlight the potential of pretrained TSFMs to enable scalable, efficient, and training-free forecasting for modern network monitoring and management systems.
\end{abstract}


\section{Introduction}

Over the past decade, paradigms such as Software-Defined Networking (SDN), Network Function Virtualization (NFV), and 5G/6G architectures have granted operators increasing programmability and control. This shift has accelerated the move toward anticipatory networking, where management decisions, such as traffic engineering, resource allocation, and service orchestration, are enacted before traffic fluctuations occur~\cite{ferreira2023forecasting}. The effectiveness of such proactive strategies depends on the accuracy of traffic forecasts, which are now central to modern network management.

Accurate forecasting supports not only capacity planning and load balancing, but also anomaly detection, congestion prevention, and quality-of-service (QoS) assurance. In anomaly detection, for example, forecasts serve as a baseline against which deviations are flagged: when the observed traffic diverges significantly from the predicted value, the event can be considered anomalous. Thus, traffic forecasting is not only a performance optimization tool but also a foundation for security and reliability in network operations and management.

Despite decades of research, several challenges persist. Network traffic is inherently dynamic, influenced by diurnal and weekly cycles, seasonal effects, user mobility, and application-level behaviors. Bursty patterns, sudden surges or drops, remain particularly hard to capture. Traditional statistical models such as the autoregressive integrated moving average (ARIMA)~\cite{box2015time, zhang2018time} and exponential smoothing models~\cite{winters1960forecasting} are lightweight but struggle with non-linearity and randomness in traffic dynamics. Deep learning approaches, including LSTMs~\cite{hochreiter1997long}, GRUs, CNNs, and Transformers~\cite{zeng2023transformers, lim2021time}, offer stronger feature extraction and outperform statistical models in many settings. However, they typically require training per dataset or even per time series, limiting their generalization and scalability. In large ISP environments, where hundreds of thousands of time series must be forecasted, this training burden becomes impractical.

Recently, time-series foundation models (TSFMs) such as Google’s TimesFM~\cite{das2024decoder}, Amazon’s Chronos~\cite{ansari2024chronos}, and IBM’s Tiny Time Mixers (TTM)~\cite{ekambaram2024tiny} have emerged as promising general-purpose forecasting solutions. Pretrained on large and diverse time-series corpora, these models demonstrate strong zero-shot and few-shot generalization across domains such as finance, energy, and transportation. However, network traffic exhibits different statistical characteristics, such as burstiness, abrupt regime shifts, and heavy-tailed volume distributions, compared to the smoother and more stationary patterns typical of financial or energy data. Consequently, despite their strong benchmark results in other domains, the effectiveness of TSFMs for computer networking remains unknown. To our knowledge, prior TSFM studies have not evaluated their applicability to network telemetry data, which exhibits unique temporal and structural dependencies. This gap motivates our investigation into whether a pretrained model like TTM, trained entirely on non-network data, can generalize effectively to ISP traffic forecasting tasks without retraining.

For evaluation, we adopt the CESNET-TimeSeries24 dataset~\cite{koumar2025cesnet}, which contains 40 weeks of real-world ISP traffic metrics across 283 institutions, 548 subnets, and over 270,000 IP addresses. The dataset includes an associated benchmark study~\cite{koumar2025comparative} that systematically evaluates strong deep learning baselines (GRU, LSTM, CNN) under standardized train/validation/test splits and forecasting horizons. Aligning with this benchmark ensures that our results are directly comparable to established baselines.

In this paper, we investigate whether TTM can outperform state-of-the-art deep learning models on ISP traffic forecasting without retraining. We evaluate both zero-shot and few-shot settings and extend the analysis with exogenous features (e.g., weekends and holidays). To structure our study, we formulate the following research questions:

\begin{itemize}
    \item \textbf{RQ1: Context sensitivity.} How does forecasting accuracy change as the context window length varies?
    \item \textbf{RQ2: Horizon scaling.} How does performance evolve as the forecasting horizon increases (from hours to days ahead)?
    \item \textbf{RQ3: Cross-level generalization.} Can TTM generalize effectively to ISP traffic across hierarchical levels (institutions, subnets, and IP addresses), without retraining?
    \item \textbf{RQ4: Few-shot adaptation.} Does limited fine-tuning provide measurable gains compared to zero-shot inference?
    \item \textbf{RQ5: Role of exogenous features.} Do auxiliary attributes (such as weekends and holidays) improve forecasting accuracy?
    \item \textbf{RQ6: Efficiency.} What are the runtime and inference costs, and are they practical for deployment in real-time ISP environments?
\end{itemize}

The main contributions of this paper are as follows:

\begin{itemize}
    \item \textbf{Evaluation of TTM on real ISP traffic.} We apply IBM’s TTM to CESNET-TimeSeries24 and evaluate its performance under the benchmark protocol. To facilitate reproducibility, we publish the source code of the evaluation on GitHub \footnote{https://github.com/FanL1u/TSFM-for-ISP-traffic-forecasting.git}.
    \item \textbf{Zero-shot state-of-the-art performance.} We show that TTM, even without fine-tuning, outperforms strong deep learning baselines (GRU, LSTM) on this dataset.
    \item \textbf{Negative result on few-shot learning.} We systematically evaluate few-shot fine-tuning and exogenous features, and find negligible improvement.
    \item \textbf{Practical implications for ISPs.} Our results demonstrate that TTM is not only accurate but also lightweight and fast, making it suitable for deployment in operational ISP environments where real-time inference and efficiency are critical.
\end{itemize}

\section{Related Work}

Network traffic prediction can be treated as a time series forecasting problem. Existing approaches broadly fall into three categories: (i) statistical models, (ii) deep learning models, and (iii) foundation models such as TSFMs and LLM-based architectures. Statistical methods assume linear temporal relationships and estimate model parameters through likelihood- or least-squares-based optimization; deep learning models (e.g., RNNs, CNNs, and Transformers) learn nonlinear temporal dependencies directly from data; and foundation models extend this idea by pretraining large neural networks on massive, heterogeneous corpora, enabling cross-domain generalization and zero/few-shot forecasting.

\subsection{Statistical Forecasting}

Early network-traffic forecasting methods relied on statistical time-series models such as exponential smoothing and the ARIMA models~\cite{hendikawati2020survey}. These approaches remain appealing for their interpretability and low computational cost, yet a linear correlation structure is assumed among the time series values, and therefore, no nonlinear patterns can be captured by the ARIMA model. The approximation of linear models to complex real-world problems is not always satisfactory. Many researchers have argued
that real-world systems are often nonlinear~\cite{zhang1998forecasting}. Hybrid extensions combining ARIMA with neural models such as ANN and LSTM~\cite{zhang2003time, khashei2011novel} have been explored to capture both linear and non-linear patterns, but their improvements often come at the expense of increased model complexity and extensive parameter tuning.

\subsection{Deep Learning for Forecasting}

Deep neural architectures have been increasingly adopted for traffic forecasting. Recurrent neural networks such as LSTMs~\cite{hochreiter1997long} and GRUs capture sequential dependencies effectively, while CNN-based models extract local temporal features. Transformer architectures have recently been adopted for their ability to model long-range dependencies and handle irregular time intervals~\cite{zeng2023transformers}. Hybrid spatio-temporal networks using Graph Neural Networks (GNNs) have achieved strong performance by modeling both temporal and spatial correlation~\cite{wu2019graph,peng2024network}. However, identifying the most suitable deep learning architecture remains challenging. Although models such as LSTMs, CNNs, and Transformers have achieved strong predictive accuracy, selecting and optimizing them often involves extensive hyperparameter tuning, with complexity increasing rapidly as model depth grows. Moreover, architectural effectiveness is highly dependent on the characteristics of the dataset, available computational resources, and the target application domain~\cite{aouedi2025deep}.

\subsection{TSFM}

Recently, TSFMs have emerged as a new paradigm for forecasting, offering strong zero-shot and few-shot generalization by pretraining on massive, heterogeneous corpora. TimesFM~\cite{das2024decoder} employs a decoder-only architecture to deliver robust cross-domain forecasting. TTM~\cite{ekambaram2024tiny} emphasizes efficiency, achieving competitive accuracy with models as small as one million parameters and real-time inference on commodity CPUs. Other meta-learning approaches, such as the Feature-Adaptive Framework (FAF)~\cite{ouyang2025faf} and transfer-learning methods~\cite{bhaumik2025ssmt}, seek to fine-tune pretrained representations efficiently with limited data.

In parallel, foundation models for wireless and mobile networking have begun to emerge. Zhang et al.\cite{chai2025uomo} introduced UoMo, a universal model for mobile-traffic forecasting, achieving strong cross-network transfer for cellular optimization tasks. Li et al.\cite{sheng2025wireless} proposed a wireless foundation model capable of multi-task prediction, including traffic forecasting, anomaly detection, and resource optimization, showing that pretrained temporal encoders can generalize across heterogeneous network traces. Beyond forecasting, Wehner et al.\cite{wehner2025exploring} explored TSFMs such as TTM for network-monitoring tasks, demonstrating accurate zero-shot prediction of Quality-of-Experience (QoE) from encrypted traffic streams while maintaining low inference latency.

\subsection{LLMs for Time-Series Forecasting}

LLM-based forecasting is emerging in the time-series domain. Time-LLM~\cite{jin2023time} reprograms frozen LLMs to accept time-series patches as textual prototypes and demonstrates competitive zero-shot and few-shot performance. Beyond generic forecasting, domain-specific adaptations are appearing: TrafficLLM uses in-context learning for wireless traffic prediction without fine-tuning~\cite{hu2024self}, and ST-LLM frames spatial-temporal traffic tokens with specialized embeddings and partially frozen attention for network traffic forecasting~\cite{liu2024spatial}. However, some critical analyses (e.g., Tan et al.~\cite{tan2024language}) argue that many LLM-based forecasting systems add computational overhead without consistent performance gains over simpler architectures. Recent work, such as TALON~\cite{sun2025adapting}, attempts to address the inherent modality gap between numeric time-series data and discrete LLM inputs by modeling temporal heterogeneity and enforcing semantic alignment.

These examples illustrate both the promise and challenge of applying LLMs to traffic forecasting, and motivate our choice to use a purpose-designed TSFM that better balances inference efficiency, domain specificity, and forecasting accuracy.

\subsection{Gap in ISP Forecasting}

While foundation models have achieved promising results in finance, energy, and transportation~\cite{fu2025financial}, their application to Internet traffic remains largely unexplored. Ferreira et al.~\cite{ferreira2023forecasting} identify generalization and transfer learning as open challenges in network-traffic prediction.

Furthermore, the majority of existing time-series foundation models, such as TimesFM~\cite{das2024decoder}, Chronos~\cite{ansari2024chronos}, and TTM~\cite{ekambaram2024tiny}, are pretrained on large-scale datasets drawn primarily from non-networking domains, including energy consumption, financial transactions, and environmental measurements. Although these datasets enable broad temporal pattern learning, it remains unclear how well such cross-domain representations transfer to Internet traffic, which exhibits distinctive characteristics such as heavy-tailed volume distributions, diurnal periodicity, and abrupt bursts linked to human activity or network events.

This raises two critical questions for the networking community: \textbf{(i) can existing foundation models pretrained on heterogeneous but non-network datasets generalize effectively to ISP traffic forecasting, and (ii) is there a need to develop networking-specific foundation models pretrained on large-scale traffic corpora?}

Our study addresses these questions by systematically evaluating TTM on the CESNET-TimeSeries24 dataset, assessing zero-shot, few-shot, and exogenous-feature settings to show whether pretrained forecasting models can match or surpass specialized deep-learning baselines while maintaining low computational overhead.

\begin{figure}[htbp]
  \centering
  \includegraphics[width=\linewidth]{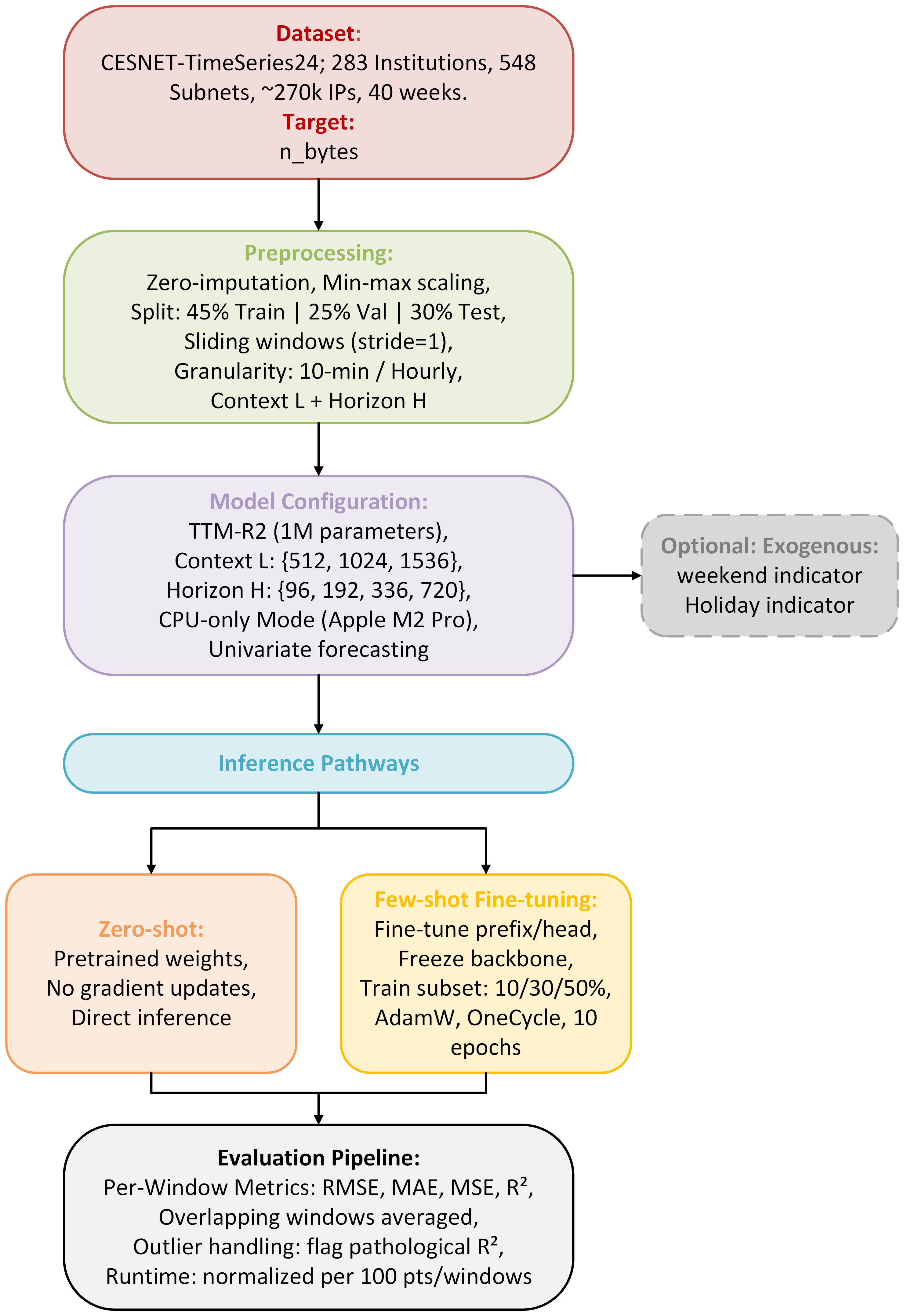}
  \caption{Overview of the proposed experimental methodology. The CESNET-TimeSeries24 dataset is preprocessed (scaling, windowing, and split into train/val/test). The TTM is used for both zero-shot inference and few-shot fine-tuning. We evaluate multiple context–horizon configurations, cross-level generalization, and the effect of exogenous features, reporting RMSE, MAE, MSE, and $R^2$ metrics.}
  \label{fig:methodology}
\end{figure}

\section{Methodology}

This section describes the experimental pipeline used to evaluate TTM-R2 (where R2 denotes the model’s release version; TTM is used throughout the paper) for ISP traffic forecasting. As illustrated in Figure~\ref{fig:methodology}, the workflow comprises five stages: (1) dataset preprocessing and window generation, (2) model configuration and context–horizon setup, (3) zero-shot and few-shot execution, (4) metric computation across overlapping windows, and (5) efficiency benchmarking on CPU-only mode. The evaluation workflow is dataset-agnostic and applies to other time-series datasets with fixed sampling intervals and sufficient historical context for context–horizon forecasting.

\subsection{Dataset and Forecasting Targets}

We evaluate our models on the CESNET-TimeSeries24 dataset~\cite{koumar2025cesnet}, a large-scale real-world ISP telemetry corpus collected from the Czech Education and Scientific Network (CESNET). The dataset spans approximately 40 weeks of traffic monitoring and contains measurements at three hierarchical aggregation levels:

\begin{itemize}
    \item Institution level: 283 organizations (universities, research centers, etc.);
    \item Subnet level: 548 institution subnets;
    \item IP level: a sampled set of $\sim$270{,}000 individual IP addresses.
\end{itemize}

Each record captures multiple traffic attributes including \textit{n\_flows}, \textit{n\_packets}, \textit{n\_bytes}, \textit{tcp\_udp\_ratio\_bytes}, and \textit{avg\_duration} etc. Data are collected at a base granularity of 10 minutes and further aggregated to hourly and daily resolution by temporal averaging, resulting in three temporal granularities:
\begin{itemize}
    \item 10-minute (10m): $\sim$40{,}000 samples per series, capturing short-term dynamics and bursty variations;
    \item Hourly (H): $\sim$6{,}700 samples per series, emphasizing diurnal and weekly periodicity.
    \item Daily (D): $\sim$280 samples per series.
\end{itemize}

\paragraph{Data characteristics}
Following the dataset’s technical validation~\cite{koumar2025cesnet}, CESNET-TimeSeries24 exhibits substantial sparsity, particularly at finer temporal resolutions. At the 10-minute aggregation level, 99.94\% of IP-level intervals contain gaps, compared to 99.21\% at 1-hour aggregation, primarily due to inactive hosts and periods of network-level silence. At higher aggregation levels, sparsity remains notable: 78\% and 86\% of intervals are missing for institutions and subnets, respectively, at 10-minute resolution, though hourly aggregation reduces these ratios to 22\% and 32\%. As expected, coarser temporal averaging mitigates missing intervals but can obscure short-lived anomalies and bursty fluctuations. This pronounced imbalance between active and inactive periods reflects a core characteristic of ISP telemetry, where most endpoints communicate intermittently following diurnal and weekly activity cycles.

\paragraph{Target selection}
Among the available metrics, we focus primarily on \textit{n\_bytes}, representing the total traffic volume per interval. We initially experimented with multiple target attributes and observed consistent performance trends across them. We therefore report \textit{n\_bytes} as the representative target for clarity and comparability with other deep-learning baseline benchmarks \cite{koumar2025comparative}. In addition, this metric directly reflects bandwidth utilization and network load, making it operationally relevant for applications such as capacity planning, congestion management, and anomaly detection in ISP networks.

\subsection{Train/Val/Test Splits}
We adopt a fixed chronological data split per series (no rolling origin, no shuffle) to maintain temporal causality and ensure reproducibility across experiments:
\begin{itemize}
    \item Train: 45\%, Validation: 25\%, Test: 30\%.
\end{itemize}
All splits are contiguous and preserve natural time ordering. A time series is excluded if any subset cannot yield at least one valid sliding window of combined length $(L + H)$, where $L$ and $H$ denote the context and prediction horizon, respectively.

\subsection{Preprocessing}
We use the \texttt{TimeSeriesPreprocessor} with min--max scaling applied independently to each time series. We fill missing values with zeros, which represent zero transmitted traffic (flows, packets, bytes, etc.). No additional interpolation or smoothing is performed beyond this zero-imputation step. Each entity (institution, subnet, or IP) is modeled as an independent univariate time series.

\paragraph{Sliding-window sampling}
Each series is segmented into overlapping samples using a stride of 1, providing exhaustive temporal coverage. For each sample, the first $L$ points form the historical context and the next $H$ points form the prediction target. This dense, stride-1 sampling ensures every possible forecasting situation within each split is evaluated, reducing sampling bias and stabilizing metrics by averaging predictions across overlapping windows.

\paragraph{Input–output tensor format}
Each sliding window maps to a single model sample. The input context tensor has shape $[L, 1]$ (or $[L, C]$ when exogenous channels are included) and the prediction target has shape $[H, 1]$, as illustrated in Figure~\ref{fig:window_example}. Among the dataset's traffic attributes, only the scaled target column is assigned to the input channel; timestamps are used solely for windowing and are not provided to the model. In batched inference the shapes become $[B, L, 1] \rightarrow [B, H, 1]$.

\begin{figure}[htbp]
\centering
\small
\begin{tabular}{c|c|c}
\toprule
\textbf{Step} & \textbf{Timestamp (hourly)} & \textbf{Scaled \textit{n\_bytes}} \\
\midrule
$t_1$ & 2024-01-15 00:00 & 0.312 \\
$t_2$ & 2024-01-15 01:00 & 0.287 \\
$\vdots$ & $\vdots$ & $\vdots$ \\
$t_L$ & 2024-02-26 07:00 & 0.445 \\
\midrule
$t_{L+1}$ & 2024-02-26 08:00 & \textit{0.461 (pred.)} \\
$\vdots$ & $\vdots$ & $\vdots$ \\
$t_{L+H}$ & 2024-03-01 07:00 & \textit{0.389 (pred.)} \\
\bottomrule
\end{tabular}
\caption{Illustration of one sliding window ($L{=}1024$, $H{=}96$, hourly). The top $L$ rows form the context input $[L, 1]$; the bottom $H$ rows are the prediction target $[H, 1]$. Values shown are after min--max scaling.}
\label{fig:window_example}
\end{figure}

\paragraph{Context–Horizon Pairs}
We restrict to model-supported $(L,H)$ pairs (Table~\ref{tab:lhpairs}).  
In the main text we standardize on hourly $H=96$ and vary $L$ for context sensitivity; for 10-minute we fix $L=1024$ and vary $H$ for horizon scaling.

\begin{table}[htbp]
\centering
\caption{Supported context ($L$) and horizon ($H$) pairs in TTM.}
\label{tab:lhpairs}
\begin{tabular}{c|c}
\toprule
Context $L$ & Prediction Horizons $H$ \\
\midrule
512 & \{48, 96, 192, 336, 720\} \\
1024 & \{96, 192, 336, 720\} \\
1536 & \{96, 192, 336, 720\} \\
\bottomrule
\end{tabular}
\end{table}

\subsection{Model and Protocols}
We compared several representative TSFMs, including TimesFM~\cite{das2024decoder}, TimeGPT~\cite{garza2023timegpt}, and TTM~\cite{ekambaram2024tiny}. TTM was selected for detailed evaluation due to its small model size (1M parameters), fast CPU inference, competitive zero-shot accuracy across public benchmarks, and open availability. The TTM variant used in our experiments strikes an effective balance between efficiency and accuracy, enabling CPU-only inference on commodity hardware while maintaining state-of-the-art forecasting performance, an essential property for scalable ISP deployments.

In our experiments, we evaluate two regimes:
\begin{itemize}
    \item Zero-shot: direct inference with pretrained weights, without any gradient updates.
    \item Few-shot: lightweight adaptation, updating prefix/head parameters while freezing the backbone.
\end{itemize}

\subsection{Few-shot Fine-tuning}
Few-shot fine-tuning evaluates whether limited adaptation can enhance task alignment while preserving TTM’s pretrained priors. We simulate data-scarce operational settings typical for newly monitored network segments by fine-tuning only on small fractions (10\%, 30\%, 50\%) of the training windows.
\begin{itemize}
    \item Optimizer: AdamW, Learning rate (LR) from built-in LR finder.
    \item Scheduler: OneCycle.
    \item Epochs: 10 max with early stopping (patience 3).
    \item Batch sizes: 64.
    \item Seed: 42.
\end{itemize}

\subsection{Exogenous Variables}
We evaluate the effect of exogenous signals by comparing runs with and without binary weekend and holiday flags. The default configuration uses univariate input only.

\subsection{Evaluation}
\paragraph{Evaluation Metrics}
Forecasts are generated for all available test windows, and evaluation metrics (RMSE, MAE, MSE, $R^2$) are computed per window and averaged across all windows within each series. This exhaustive averaging approximates a continuous online forecasting process and mitigates stochastic variance due to bursty traffic intervals.

We report RMSE and $R^2$ in the main text (MAE and MSE are omitted for brevity), as they jointly capture both magnitude error and explained variance, complementary indicators of forecasting accuracy. Timing is normalized per 100 points and per 100 windows:

\begin{itemize}
    \item Eval time: from HuggingFace \texttt{Trainer}’s \texttt{eval\_runtime}.
    \item Train time: from \texttt{train\_runtime} (few-shot only).
\end{itemize}

To control for cold-start overhead, the first evaluated series is excluded when aggregating runtime.

\paragraph{Outlier Handling}
We retain all metrics but flag pathological $R^2$ (e.g., highly negative values from nearly-constant targets). RMSE/MAE remain valid. For visualization, we use box plots with outliers shown as points.

\subsection{Hardware}
All experiments were executed on a single commodity machine, MacBook Pro (Apple M2 Pro, 16 GB RAM) running macOS 14.7.1 (23H222), using CPU-only computation (no GPU or MPS acceleration). This configuration tests the practical deployability of foundation models under computation resource constraints and ensures full reproducibility without distributed or accelerator-specific optimizations.

\subsection{Experiment Matrix}
Table~\ref{tab:lhpairs} summarizes the corresponding context–horizon configurations used in these experiments. The main paper reports:

\begin{itemize}
    \item Hourly (H=96): context sensitivity ($L\in\{512,1024,1536\}$), cross-level robustness (Institutions, Subnets, IP sample), few-shot curves (10\%, 30\%, 50\%), exogenous toggle (on/off).
    \item 10-minute: horizon scaling ($H\in\{96,336,720\}$, $L=1024$), context sensitivity (L sweep at $H=336$), exogenous toggle.
\end{itemize}

\section{Results and Discussion}

\begin{figure}[htbp]
    \centering
    \includegraphics[width=0.85\linewidth]{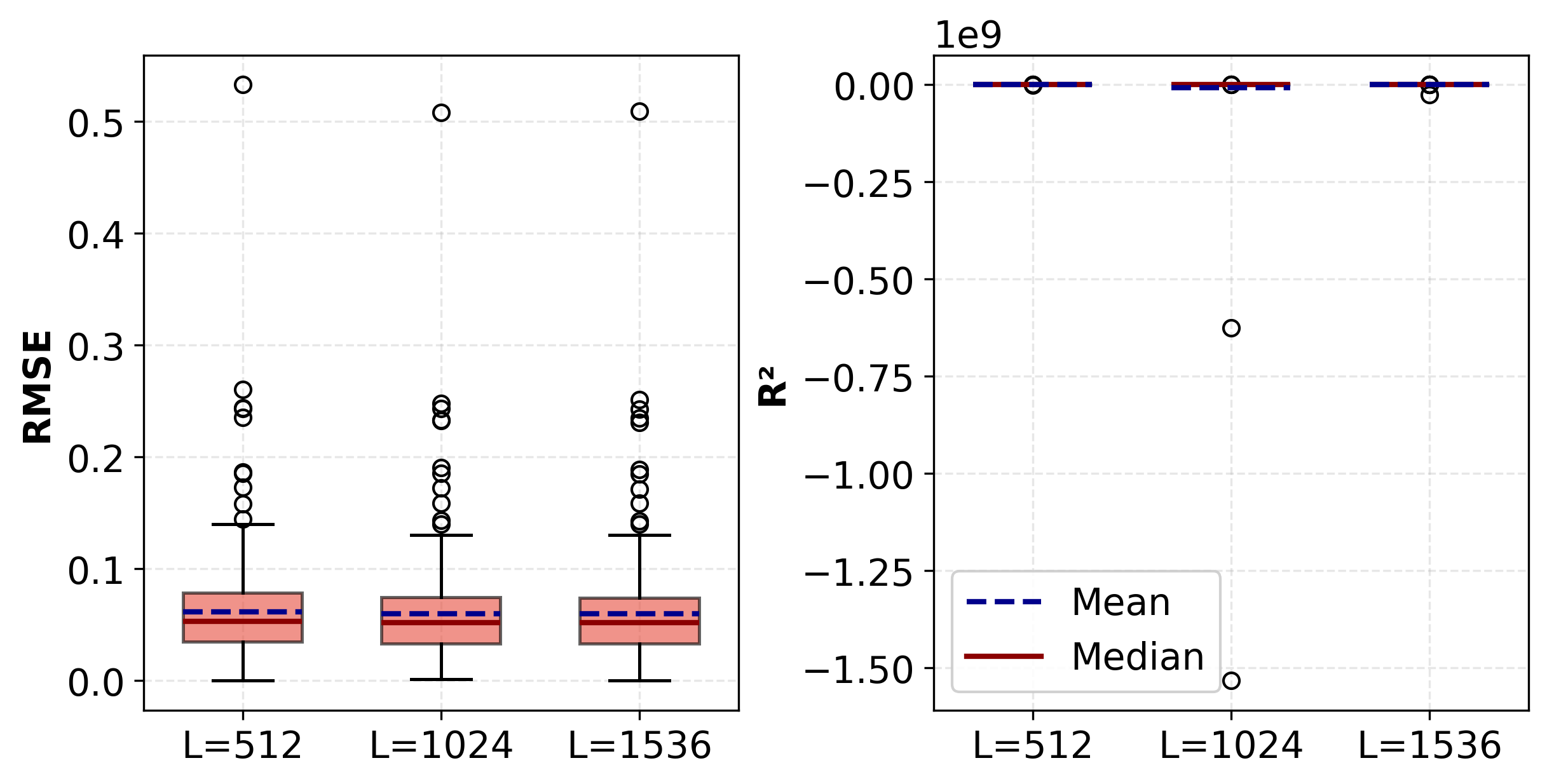}
    \caption{Hourly: Context sensitivity (Institutions, $H=96$).}
    \label{fig:hourly_context_sensitivity}
\end{figure}

\subsection{Effect of Context Length (RQ1)}

Figure~\ref{fig:hourly_context_sensitivity} and Figure~\ref{fig:10min_context_sensitivity} (shown in the Appendix) illustrate how varying the context length ($L$) influences forecasting performance across hourly and 10-minute aggregation levels. As both exhibit similar trends, only the hourly results are discussed here due to space constraints.

For hourly forecasts ($H=96$), the RMSE remains nearly unchanged as $L$ increases from 512 to 1536, indicating that the model already encodes sufficient temporal memory with moderate context. Similarly, at 10-minute resolution ($H=336$), longer lookback windows show no consistent improvement in either RMSE or $R^2$. These results suggest that TTM’s pretrained temporal representations capture recurring diurnal and weekly cycles even when short historical windows are used.

\begin{figure}[htbp]
    \centering
    \includegraphics[width=0.85\linewidth]{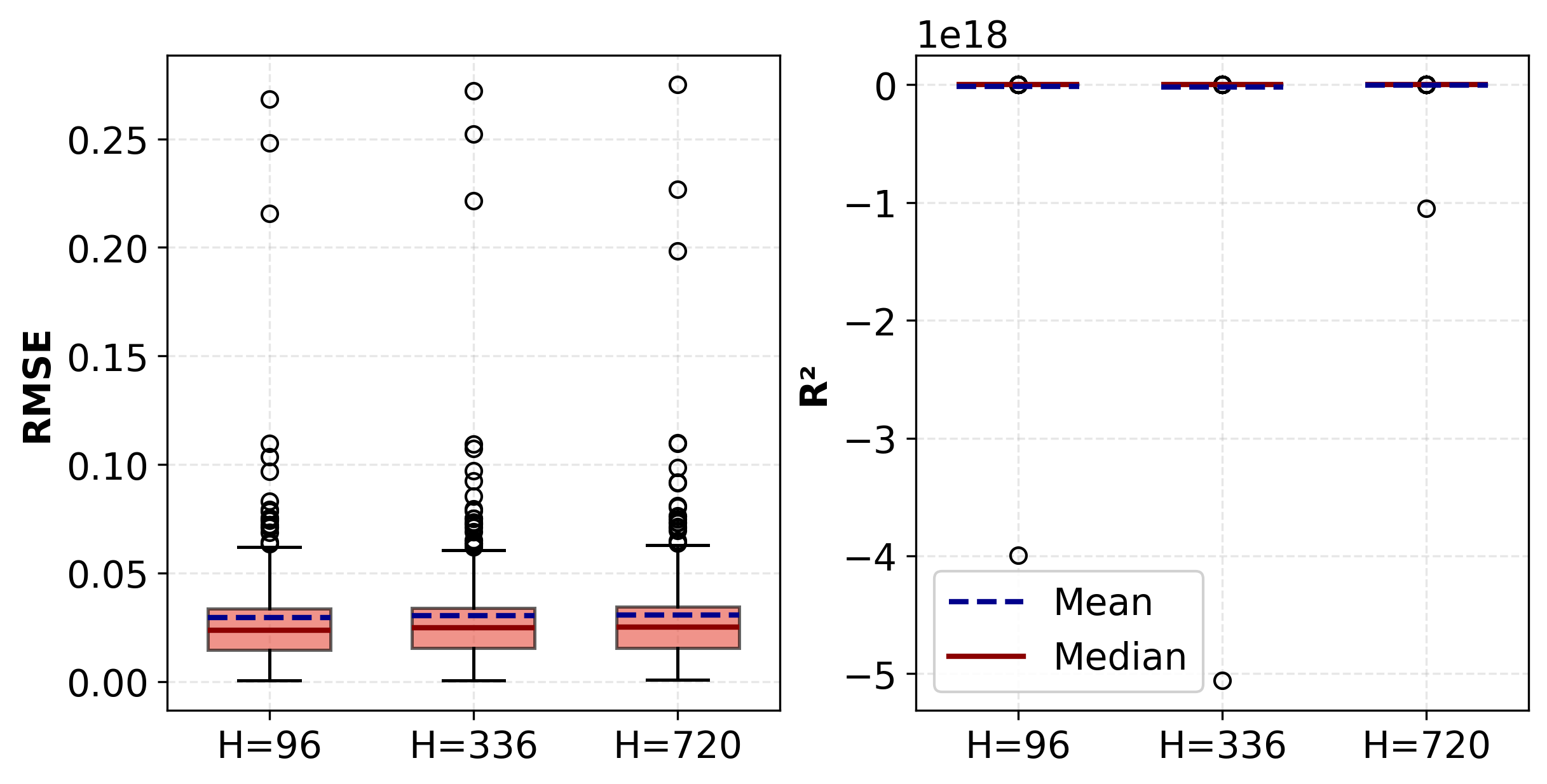}
    \caption{10-minute: Horizon scaling ($L=1024$).}
    \label{fig:10min_horizon_scaling}
\end{figure}

\subsection{Effect of Forecast Horizon (RQ2)}

Figure~\ref{fig:10min_horizon_scaling} evaluates the effect of extending the prediction horizon ($H$) from 96 to 720 with a fixed context length ($L=1024$). Across all horizons, TTM maintains low RMSE and stable $R^2$ values, demonstrating robust long-term forecasting capability. The modest degradation observed at $H=720$ reflects expected uncertainty propagation as the forecast extends beyond one day.

In contrast to this stability, the CESNET benchmark by Koumar et al.~\cite{koumar2025cesnet} reported pronounced degradation in traditional deep learning models (GRU, LSTM, CNN) as prediction windows lengthened. They observed that forecasting the immediate next value yielded substantially higher accuracy than multi-step horizons, attributing the degradation to cumulative error. Even the most stable architectures, LSTM and GRU, showed significant performance drops as horizons increased from 1 to 168 hours, emphasizing the difficulty of maintaining consistency over extended prediction intervals.

TTM, however, exhibits robustness across the tested horizons ($H \in \{96, 336, 720\}$) at 10-minute resolution. Figure~\ref{fig:10min_horizon_scaling} shows nearly constant RMSE and $R^2$ across horizons, despite operating in a \textit{zero-shot} configuration without task-specific retraining. This stands in contrast to the fully trained deep learning baselines in~\cite{koumar2025cesnet}, which required dataset-specific optimization yet still exhibited rapid error accumulation with longer horizons.

Interestingly, Koumar et al.~\cite{koumar2025cesnet} also noted that longer prediction windows can be operationally advantageous, reducing inference frequency and smoothing transient fluctuations, potentially improving anomaly-detection sensitivity. TTM’s ability to sustain accuracy at large horizons while maintaining sub-0.5\,s evaluation time per 100 points directly addresses this need. Unlike traditional models that trade off accuracy for temporal range, TTM provides reliable multi-hour to multi-day forecasts with minimal computational overhead, validating the potential of pretrained foundation models for proactive and scalable ISP traffic management.

\begin{table*}[htbp]
\centering
\caption{Forecasting Accuracy Summary Across All Experimental Configurations. Values reported as mean ± standard deviation after removing top/bottom 5\% outliers.}
\label{tab:accuracy_summary}
\begin{tabular}{llccp{3cm}p{3cm}}
\toprule
\textbf{Experiment} & \textbf{Context L} & \textbf{Horizon H} & \textbf{Few-shot \%} & \textbf{RMSE} & \textbf{R²} \\
\midrule
\multicolumn{6}{l}{\textit{Context Length Sensitivity (Hourly, Institutions)}} \\
Hourly L=512 & 512 & 96 & 0 & 0.0566 ± 0.0269 & 0.2127 ± 0.1388 \\
Hourly L=1024 & 1024 & 96 & 0 & 0.0551 ± 0.0263 & 0.2460 ± 0.1571 \\
Hourly L=1536 & 1536 & 96 & 0 & 0.0551 ± 0.0265 & 0.2498 ± 0.1576 \\
\midrule
\multicolumn{6}{l}{\textit{Cross-Level Generalization (Hourly, L=1024, H=96)}} \\
Cross: Institutions & 1024 & 96 & 0 & 0.0551 ± 0.0263 & 0.2460 ± 0.1571 \\
Cross: Subnets & 1024 & 96 & 0 & 0.0541 ± 0.0365 & 0.1685 ± 0.1672 \\
Cross: IPs & 1024 & 96 & 0 & 0.0317 ± 0.0262 & $-1.3872 \pm 8.6764$ \\
\midrule
\multicolumn{6}{l}{\textit{Horizon Scaling (10-minute, Institutions, L=1024)}} \\
10-min H=96 & 1024 & 96 & 0 & 0.0259 ± 0.0142 & 0.0975 ± 0.0982 \\
10-min H=336 & 1024 & 336 & 0 & 0.0268 ± 0.0144 & 0.0654 ± 0.0725 \\
10-min H=720 & 1024 & 720 & 0 & 0.0271 ± 0.0146 & 0.0456 ± 0.0638 \\
\midrule
\multicolumn{6}{l}{\textit{Few-Shot Learning (Hourly, Institutions, L=1024, H=96)}} \\
Few-shot: 0\% (zero-shot) & 1024 & 96 & 0 & 0.0551 ± 0.0263 & 0.2460 ± 0.1571 \\
Few-shot: 10\% & 1024 & 96 & 10 & 0.0718 ± 0.0380 & $-0.2170 \pm 0.3600$ \\
Few-shot: 30\% & 1024 & 96 & 30 & 0.0605 ± 0.0299 & 0.1196 ± 0.1847 \\
Few-shot: 50\% & 1024 & 96 & 50 & 0.0622 ± 0.0263 & 0.1936 ± 0.1503 \\
\bottomrule
\end{tabular}
\end{table*}

\begin{figure}[htbp]
    \centering
    \includegraphics[width=0.85\linewidth]{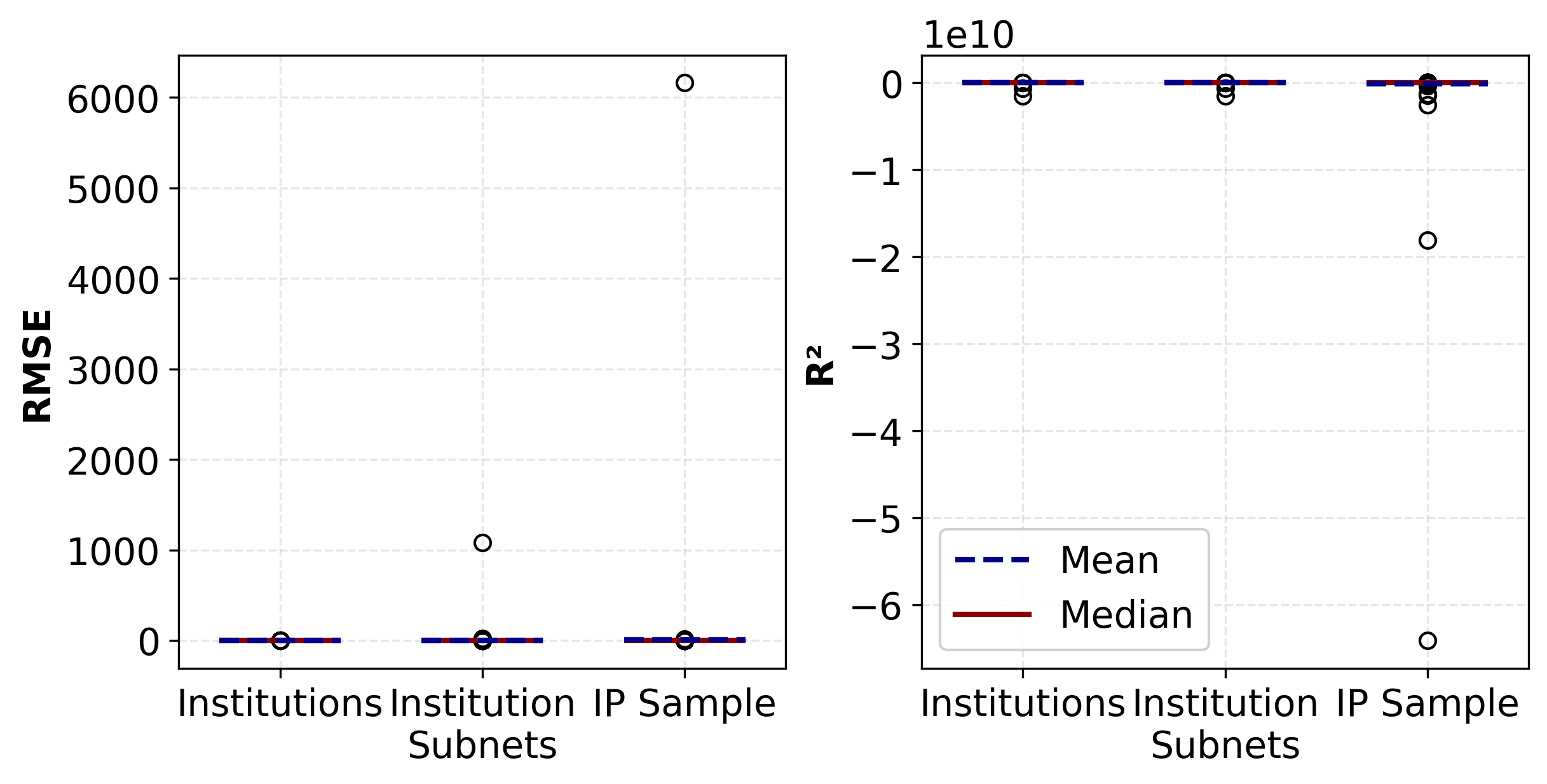}
    \caption{Hourly: Cross-level generalization (Institutions $\rightarrow$ Subnets/IPs, $L=1024$, $H=96$).}
    \label{fig:hourly_cross_level}
\end{figure}

\begin{table*}[htbp]
\centering
\caption{Inference and Training Efficiency Across Experimental Configurations (CPU-only, Apple M2 Pro, 16GB RAM). All values reported as mean ± standard deviation. Training time only applicable to few-shot experiments.}
\label{tab:efficiency}
\begin{tabular}{lccccc}
\toprule
\textbf{Configuration} & 
\textbf{Eval Time} & 
\textbf{Eval Time} & 
\textbf{Throughput} & 
\textbf{Train Time} \\
& \textbf{per 100 pts (s)} & 
\textbf{per 100 win (s)} & 
\textbf{(samples/s)} & 
\textbf{per 100 pts (s)} \\
\midrule
\multicolumn{5}{l}{\textit{Context Length Sensitivity (Hourly, Institutions, H=96)}} \\
L=512   & 0.040 ± 0.002 & 0.042 ± 0.002 & 2366 ± 113 & — \\
L=1024  & 0.044 ± 0.003 & 0.046 ± 0.003 & 2191 ± 131 & — \\
L=1536  & 0.046 ± 0.004 & 0.049 ± 0.004 & 2065 ± 152 & — \\
\midrule
\multicolumn{5}{l}{\textit{Cross-Level Generalization (Hourly, L=1024, H=96)}} \\
Institutions  & 0.044 ± 0.003 & 0.046 ± 0.003 & 2191 ± 131 & — \\
Subnets       & 0.052 ± 0.006 & 0.054 ± 0.006 & 1856 ± 189 & — \\
IP Sample     & 0.096 ± 0.028 & 0.101 ± 0.030 & 1081 ± 337 & — \\
\midrule
\multicolumn{5}{l}{\textit{Horizon Scaling (10-minute, Institutions, L=1024)}} \\
H=96   & 0.376 ± 0.042 & 0.379 ± 0.042 & 268 ± 37 & — \\
H=336  & 0.317 ± 0.027 & 0.326 ± 0.028 & 310 ± 36 & — \\
H=720  & 0.292 ± 0.033 & 0.310 ± 0.035 & 328 ± 51 & — \\
\midrule
\multicolumn{5}{l}{\textit{Few-Shot Fine-Tuning (Hourly, Institutions, L=1024, H=96)}} \\
10\% data  & 0.442 ± 0.009 & 0.464 ± 0.010 & 216 ± 4  & 4.21 ± 1.28 \\
30\% data  & 0.448 ± 0.014 & 0.470 ± 0.015 & 213 ± 6  & 5.28 ± 1.09 \\
50\% data  & 0.445 ± 0.012 & 0.467 ± 0.012 & 214 ± 5  & 5.48 ± 0.91 \\
\bottomrule
\end{tabular}
\end{table*}

\subsection{Cross-Level Generalization (RQ3)}

Figure~\ref{fig:hourly_cross_level} evaluates TTM’s ability to generalize across hierarchical aggregation levels, institutions, subnets, and individual IPs, without retraining. 
The RMSE and $R^2$ distributions remain comparable across levels, indicating that TTM effectively captures structural patterns that transfer between spatial granularities. 
This robustness highlights the model’s potential to serve as a unified forecasting backbone across diverse monitoring scopes.

In the CESNET benchmark, Koumar et al.~\cite{koumar2025cesnet} reported that “the forecasting precision of deep learning models is strongly influenced by the level of aggregation.” 
They observed that performance was highest at the institution level, lower at the subnet level, and weakest for individual IP addresses, an effect driven by increasing data sparsity and irregularity at finer granularities. 
Their correlation analysis further revealed a strong negative relationship ($r=-0.69$) between the missing-data ratio and $R^2$ at the IP level, confirming that fine-grained traffic traces are inherently more volatile and underdetermined.

Our findings with TTM align with this general trend but demonstrate greater resilience. 
As summarized in Table~\ref{tab:accuracy_summary}, mean $R^2$ values remain positive and stable for institutions ($\sim$0.25) and subnets ($\sim$0.17), while declining at the IP level ($\sim$–1.39) due to extreme sparsity. 
Unlike the baseline study, which suggested that distinct modeling strategies may be needed at each aggregation level (e.g., recurrent models for institutions, specialized preprocessing for IPs), TTM maintains a single unified architecture across all levels, without any retraining or architectural modification.

This result underscores one of the key advantages of time-series foundation models: \textbf{a “train-once, deploy-everywhere” paradigm}, where a single pretrained model can forecast at multiple granularities with minimal degradation. 
Such cross-level robustness is particularly valuable in ISP environments containing hundreds of thousands of monitored entities, where training and maintaining specialized models per hierarchy would be computationally infeasible. TTM thus provides a scalable and operationally efficient alternative to traditional deep learning pipelines while preserving deployment simplicity.

\begin{figure}[htbp]
    \centering
    \includegraphics[width=0.95\linewidth]{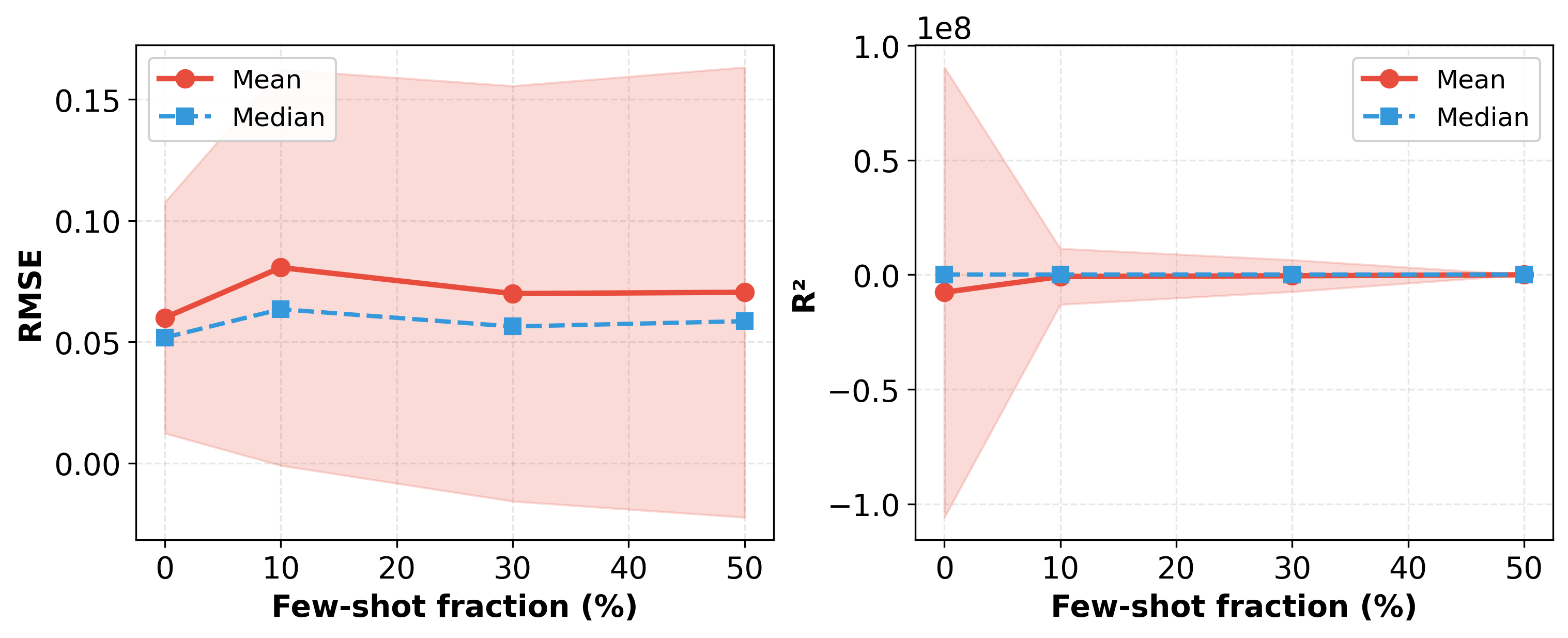}
    \caption{Hourly: Few-shot learning curve (Institutions, $L=1024$, $H=96$). Mean and median RMSE/$R^2$ shown across 10\%, 30\%, and 50\% fine-tuning fractions.}
    \label{fig:hourly_fewshot_curve}
\end{figure}

\begin{figure}[htbp]
    \centering
    \includegraphics[width=0.85\linewidth]{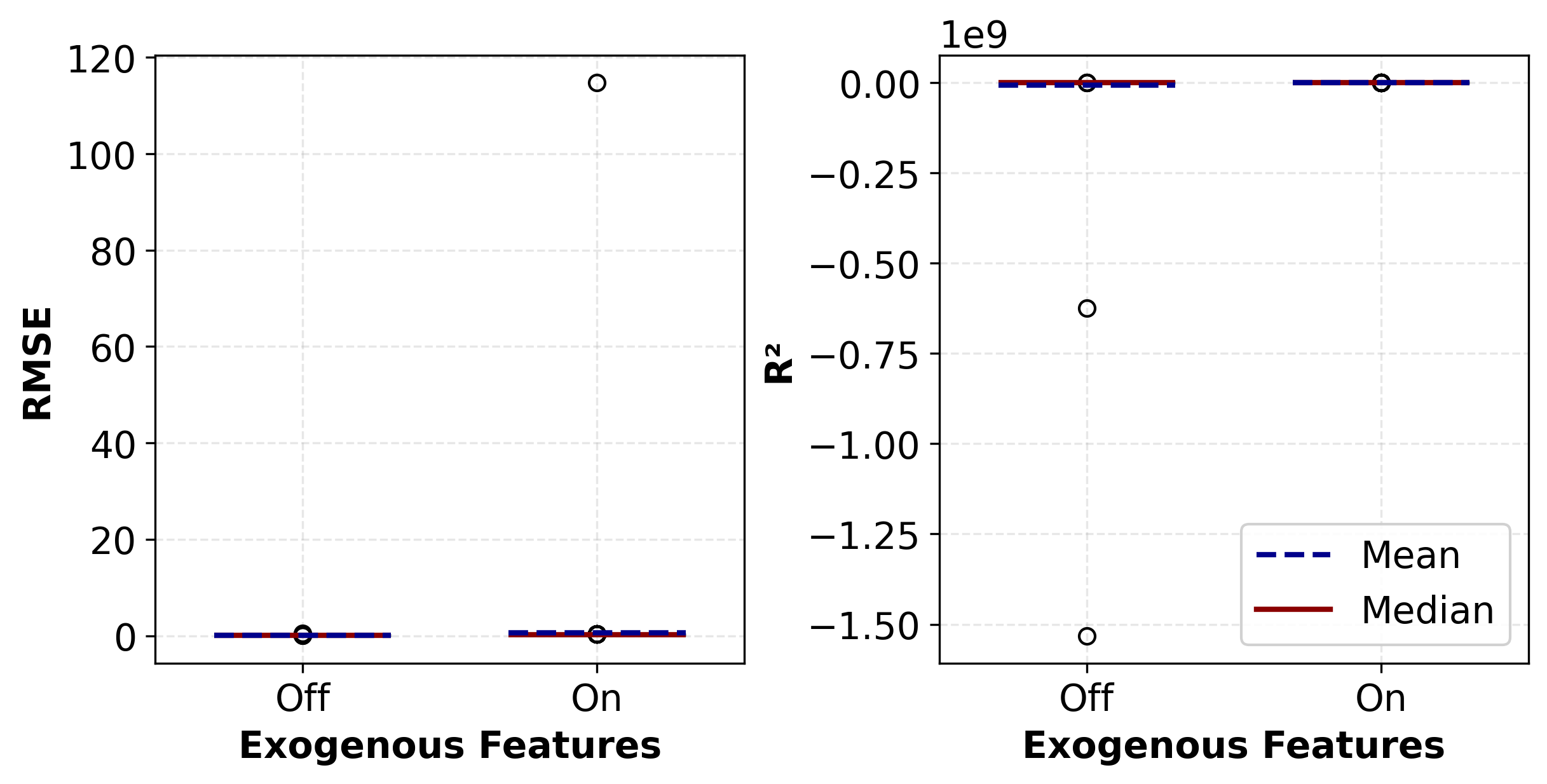}
    \caption{Hourly: Exogenous feature toggle (Institutions, $L=1024$, $H=96$).}
    \label{fig:hourly_exogenous_toggle}
\end{figure}

\subsection{Few-Shot Adaptation and Exogenous Variables (RQ4, RQ5)}

We further assess TTM’s adaptability through two complementary experiments: few-shot fine-tuning and exogenous-variable augmentation.

Figure~\ref{fig:hourly_fewshot_curve} shows that as the fraction of fine-tuning data increases from 10\% to 50\% ($L=1024$, $H=96$), RMSE improves modestly (from $\sim$0.075 to $\sim$0.070), while $R^2$ remains largely stable. This indicates that limited adaptation can refine magnitude estimation but does not substantially alter the model’s overall calibration.

In contrast, Figure~\ref{fig:hourly_exogenous_toggle} compares runs with and without auxiliary weekend and holiday indicators. The inclusion of these exogenous features yields negligible performance gains, suggesting that TTM already encodes robust periodic representations directly from intrinsic temporal patterns.

Together, these results highlight TTM’s robustness and practicality: minimal fine-tuning or auxiliary data are needed for accurate forecasting, simplifying deployment in real-world ISP settings where labeled data and external context are often limited.

\subsection{Deployability and Efficiency (RQ6)}

Computational efficiency is a key requirement for operational ISP deployment. Unlike conventional deep learning models that rely on GPUs and extensive retraining, TTM delivers strong forecasting accuracy entirely on CPU hardware with minimal overhead.
All experiments were executed on a MacBook Pro (M2 Pro, 16 GB RAM, CPU-only).
Table~\ref{tab:efficiency} summarizes inference and training costs across configurations. For hourly forecasting at the institutional level ($L=1024$, $H=96$), TTM achieves 0.044 s per 100 points ($\sim$2,200 samples/s). Efficiency remains stable across context lengths ($L=512$–1536) with only a 0.006 s difference.

Inference cost varies modestly by aggregation level, 0.044 s for institutions, 0.052 s for subnets, and 0.096 s (±0.028) for IPs, reflecting data sparsity. Even at the IP level, throughput exceeds 1,000 samples/s, sufficient for large-scale monitoring.

At 10-minute resolution, evaluation time decreases from 0.376 s ($H=96$) to 0.292 s ($H=720$) as longer horizons amortize encoding cost across more outputs. Throughput consistently exceeds 260 samples/s, enabling near-real-time forecasting.

Few-shot adaptation adds roughly a 10× overhead (0.445 s vs. 0.044 s) and 4–5.5 s training time. Zero-shot performance therefore provides a practical default for large ISP workloads.

Compared with deep-learning baselines from the CESNET benchmark~\cite{koumar2025cesnet}, TTM is 1.5× faster than the best GRU-FCN model (0.044 s vs. 0.065 s) and matches subnet-level latency (0.052 s). Models such as GRU, LSTM, and InceptionTime required 0.08–0.15 s, while ResNet and RCLSTM exceeded 0.3 s. Crucially, TTM achieves this performance in zero-shot mode without task-specific training, demonstrating that foundation models can deliver both computational efficiency and operational simplicity.

\subsection{Quantitative Summary}

Table~\ref{tab:accuracy_summary} reports the aggregated metrics (mean ± std across all evaluated series). TTM maintains consistent accuracy across context lengths and horizons, with RMSE around 0.025–0.056 and stable inference times (Table \ref{tab:efficiency}) even on CPU-only mode. R² remains positive for institutional and subnet levels, with degradation at the IP level due to flat or zero-variance series. These results confirm the model’s robustness and computational efficiency across traffic hierarchies.

\subsection{Limitations and Future Work}

Despite encouraging results, several limitations remain.  

First, this study focuses on univariate forecasting of \textit{n\_bytes}, which, while operationally meaningful, does not capture cross-metric dependencies such as the joint dynamics between \textit{flows}, \textit{packets}, and \textit{duration}. Extending TTM evaluation to multivariate forecasting will clarify whether pretrained models can internalize network-level correlations.

Second, the CESNET-TimeSeries24 dataset lacks labeled anomalies, constraining evaluation to forecasting rather than detection. Future work can explore using prediction residuals as unsupervised anomaly indicators and testing the same model on labeled intrusion datasets.

Third, while this work evaluated traditional few-shot fine-tuning, recent advances such as Google’s TimesFM demonstrate that in-context few-shot learning~\cite{das2024context}, conditioning the model on example sequences at inference time, can achieve performance comparable to fine-tuning without any gradient updates. Although TTM’s mixer-based architecture does not natively support in-context adaptation, exploring analogous strategies (e.g., contextual embedding prompts or dynamic temporal adapters) may yield similar benefits.

Fourth, IP-level performance naturally degrades under extreme sparsity and zero inflation. Several directions could mitigate this limitation in future work. First, temporal aggregation (e.g., daily or weekly rollups) can reduce intermittency at the cost of temporal resolution. Second, two-stage intermittent forecasting, predicting activity occurrence followed by conditional magnitude, provides a principled way to handle zero-inflated traffic series. Third, context augmentation with history-based features (e.g., rolling activity counts, time since last non-zero event) and fine-grained calendar signals may remain informative even during prolonged zero intervals. Finally, hierarchical information sharing, where subnet- or institution-level signals regularize IP-level predictions, could allow fine-grained forecasts to borrow strength from more stable aggregates.

\section{Conclusion}
We presented a systematic evaluation of TTM for ISP traffic forecasting on the CESNET-TimeSeries24 dataset. Across temporal resolutions, horizons, and aggregation levels, TTM delivers robust \emph{zero-shot} performance with low RMSE and stable $R^2$, while running efficiently on CPU-only hardware (e.g., $\approx$0.04–0.05\,s per 100 points at hourly resolution and $\approx$0.29–0.38\,s per 100 points at 10-minute resolution). Accuracy is largely insensitive to context length, and the model generalizes from institutions to subnets; performance naturally degrades at the IP level due to sparsity and intermittency.

Importantly, few-shot fine-tuning (10–50\% of train windows) did not materially improve accuracy under our settings: mean RMSE changed only marginally, median RMSE/$R^2$ were effectively unchanged, and the procedure incurred additional cost (about $10\times$ higher per-100-point runtime during fine-tuning and $\sim$4–6\,s per 100 points of train time). These results suggest that zero-shot TTM is already competitive for ISP forecasting.

Overall, our results answer RQ1--RQ6 by showing that TTM is largely insensitive to context length, remains robust as the forecast horizon increases, generalizes across aggregation levels, gains little from few-shot adaptation or simple exogenous features, and delivers practical CPU-only inference efficiency for ISP-scale deployment.

\bibliographystyle{IEEEtran}
\bibliography{references}

\appendices

\section{Additional Results}
This appendix provides supplementary figures and tables supporting the main analysis.

\begin{figure}[htbp]
    \centering
    \includegraphics[width=0.85\linewidth]{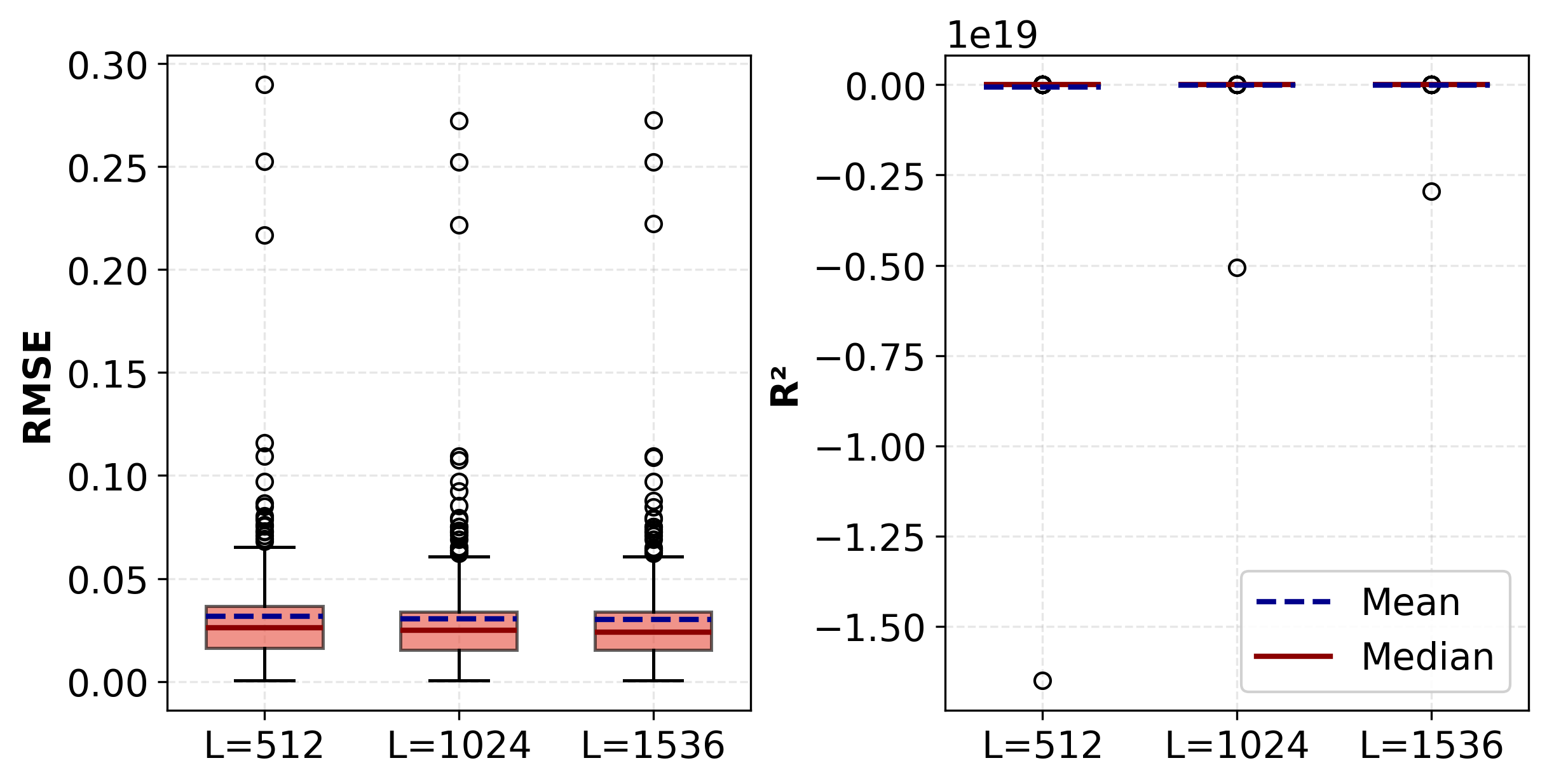}
    \caption{10-minute: Context sensitivity (Institutions, $H=336$).}
    \label{fig:10min_context_sensitivity}
\end{figure}

\end{document}